\documentclass[amsmath,amssymb,nofootinbib]{revtex4}
\usepackage{epsfig}
\usepackage{adjustbox}

\begin{document}
\title{Parameterizing the SFC Baryogenesis Model}
% Dependence of the Generated Baryon Charge on the Parameters of the SFC
%Baryogenesis Model}
\author{Daniela Kirilova$^{1,2}$, Mariana Panayotova$^{1}$}
\affiliation{$^{\it 1}$Institute of Astronomy at the Bulgarian
Academy of Sciences, Sofia, Bulgaria\\
$^{\it 2}$ Joint Institute for Nuclear Research, Dubna,
Russia\\dani@astro.bas.bg, mariana@astro.bas.bg}

%----------------------------------------------------------------------------------------
%   ABSTRACT
%----------------------------------------------------------------------------------------

\begin{abstract}

\noindent We have numerically explored the Scalar Field Condensate
baryogenesis model for numerous sets of model's parameters, within
their natural range of values. We have investigated the evolution of
the baryon charge carrying field, the evolution of the baryon charge
contained in the scalar field condensate
 and the final value of the generated baryon charge on the
model's parameters: the  gauge coupling constant $\alpha$, the
Hubble constant at the inflationary stage $H_I$, the mass $m$, the
self-coupling constants $\lambda_i$.
\end{abstract}

\maketitle % Insert title

%----------------------------------------------------------------------------------------
%   ARTICLE CONTENTS
%----------------------------------------------------------------------------------------

%\begin{multicols}{2} % Two-column layout throughout the main article text

\section*{Introduction}

%\lettrine[nindent=0em,lines=3]{L}

There exists baryon asymmetry $\beta \ne 0$ in the neighborhood of
our Galaxy, within 20 Mpc. $\beta$ is usually parameterized as
$\beta=(n_b-n_{\bar{b}})/n_{\gamma}$, where $n_b$ is the number
density of baryons, $n_{\bar{b}}$ of antibaryons and  $n_{\gamma}$
is the number density of photons. Contemporary {\bf observational
knowledge on the baryon density of the Universe} is based mainly on
the following sets of precise observational data: data based on Big
Bang Nucleosynthesis (BBN), i.e. the determination of the baryon
density from the requirement of consistency between theoretically
predicted and observationally measured abundances of the
primordially produced light elements \cite{beri12}; measurements of
Deuterium towards
 low metallicity distant quasars compared with BBN predicted D
\cite{pett12}; CMB anisotropy measurements (see WMAP
\cite{bennett12} and Planck \cite{planck13}), allowing precise
determination of the main Universe characteristics, including the
baryon density.

Namely, the consistency between theoretically obtained and
observationally measured abundances of the light elements produced
in BBN  \cite{beri12} requires that the baryon-to-photon density is
in the range:

\begin{equation} 5.1 \times 10^{-10} \le (n_b/n_\gamma)_{BBN} \le 6.5 \times 10^{-10}
\quad at \quad 95\% \quad CL
\end{equation}

The information from measurements of Deuterium towards low
metallicity quasars combined with BBN data \cite{pett12} points to:

\begin{equation} (n_b/n_{\gamma})_D = 6 \pm 0.3 \times 10^{-10} \quad at \quad 95\% \quad CL
\end{equation}

The most precise determination is provided by the measurements of
the CMB anisotropy ($z \sim 1000$). Recent results by WMAP9
\cite{bennett12} point to:

\begin{equation} (n_b/n_{\gamma})_{WMAP} = 6.19 \pm 0.14 \times 10^{-10}
\quad at \quad 68\% \quad CL
\end{equation}

The up-to-date data from Planck project \cite{planck13} point to:

\begin{equation} (n_b/n_{\gamma})_{Planck} = 6.108 \pm 0.038 \times 10^{-10}
\quad at \quad 68\% \quad CL
\end{equation}

{\bf Observational constraints exist on the presence of the
antimatter} in our local vicinity (within tens of Mpc), mainly based
on Cosmic Ray data~\cite{CR1, CR0,CR2,CR3, CR4} and Gamma Ray
data~\cite{gamma1,gamma2,gamma3,gamma4,gamma5}. No antimatter in
astronomically considerable amounts has been
observed/detected.~\footnote{Still, domains of antimatter are not
absolutely ruled out~\cite{anti1, anti2,rev2}.} Hence, the baryon
asymmetry inn our neighborhood is $\beta \sim n_b/n_{\gamma}$.
Observational evidence for matter-antimatter asymmetry in the
Universe has been recently reviewed in refs.\cite{rev1,rev2}.

In case  this  locally observed asymmetry is a {\it global}
characteristic of the Universe, i.e. baryon asymmetry of the
Universe (BAU), it may be due to the generation of a baryon excess
at some early stage of the Universe that, eventually diluted during
its further evolution, determined the value observed today.

A. Sakharov~\cite{sakh67} defined the conditions for the generation
of predominance of matter over antimatter from initially symmetric
state of the early Universe.  Namely, these are: non-conservation of
baryons, C and CP-violation and deviation from thermal equilibrium.
None of these conditions is obligatory. Different baryogenesis
scenarios, where some of the Sakharov's requirements are not
fulfilled have been discussed in ref.~\cite{dol92}.

 The exact (Nature chosen)  baryogenesis mechanism is not known
yet. It is known that baryon asymmetry may not be postulated as an
initial condition, in case of inflationary early stage of the
Universe evolution,  and it should have been generated in the period
after inflation and before BBN epoch \cite{dolgzelsaz88}.

   {\bf Numerous baryogenesis scenarios exist} today,
which aim is to explain the observed baryon asymmetry, its sign and
its value. For a review see refs.~\cite{dine,buchm,buch,dolgzel81}.
In ~\cite{buch} different mechanisms for generation both of the
baryon asymmetry and the dark matter of the Universe and their
dependence on the reheating temperature have been discussed. The
most studied among the baryogenesis scenarios are  Grand Unified
Theories (GUT) baryogenesis \cite{sakh67}, Electroweak (EW)
baryogenesis \cite{kuz85,EU1,EU2,EU3,EU4},
Baryogenesis-through-leptogenesis (often called leptogenesis)
\cite{fuku86,asaka1, asaka2}, Affleck-Dine (AD) baryogenesis
\cite{affdine85}, etc.

{\bf GUT baryogenesis} is the earliest baryogenesis scenario,
proceeding at GUT unification scale $M_{GUT}$. However, most
inflationary models predict reheating temperature below this scale.
Besides, successful unification requires supersymmetry.  SUSY
implies the existence of gravitinos, which are too numerously
produced unless the reheating temperature is well below
$M_{GUT}$~\cite{sazin,weinb}.

{\bf EW baryogenesis} is   theoretically attractive because it
relies only upon weak scale physics and is experimentally testable
scenario.
 For a review see~\cite{EUrev,cline,EU,EU2}. However, the simplest
and most appealing version of this scenario cannot generate within
the Standard Model the observed value of the baryon asymmetry,
because of the insufficient CP-violation~\cite{CP1,CP2,CP3} induced
by CKM phase and the requirement of first order electro-weak
transition, possible only for Higgs boson mass considerably smaller
than the detected one by ATLAS and CMS collaborations.

EW baryogenesis in Minimal Supersymmetric Standard Model was
considered~\cite{huet}. Now MSSM window for EW baryogenesis is
substantially narrowed  by the
 experimental data from the Large Hadron Collider (LHC)~\cite{MSSMe,MSSMe2,MSSMe3} and recent electron dipole measurements. Constraints on EW
baryogenesis in case of a minimal extension of the Standard Model
from current data from LHC have been discussed as
well~\cite{damgaard}. The
 viable parameter space is considerably reduced. Baryogenesis in next-to minimal SM
 is being discussed now~\cite{EUrev2}.

{\bf Baryogenesis-through-leptogenesis} is a plausible possibility.
Baryon asymmetry  in this scenario is  created  before  the
electroweak phase transition, which then gets converted to  the
baryon asymmetry  in the  presence  of  (B+L) violating  anomalous
processes. For a review see ref.~\cite{buchm,david,riotto,Lr}. It
has become especially attractive after the discovery of non-zero
neutrino masses in neutrino oscillations experiments. Baryogenesis
through leptogenesis mechanisms in different extensions of the SM
are studied.

Possibilities for falsifying  concrete realizations of high scale
leptogenesis from recent LHC data have been
proposed~\cite{frere,Le}.

{\it Neutrino Minimal Standard Model} (NMSM) can potentially account
simultaneously for baryon and dark matter generation and neutrino
oscillations~\cite{NMSM}. NMSM is testable at colliders and in
astrophysical observations. For a recent review and constraints from
collider experiments, astrophysics and cosmology see
ref.~\cite{NMSMr}.

{\bf AD baryogenesis} scenario~\cite{affdine85,dine} is one of the
most promising today baryogenesis scenarios, compatible with
inflation. Nice reviews of AD baryogenesis contemporary status can
be found in refs.~\cite{dine,mazumdar}. AD baryogenesis has numerous
attractive features. A short list of these is as follows: (i)  It is
extremely efficient - it can produce equal or much bigger than the
observed baryon asymmetry; (ii) It can be realized at lower energy,
i.e. relatively late in the Universe evolution. I.e. it is
consistent with the low energy scales after inflation; (iii) AD
condensate can be generated generically in different cosmological
models; (iv) It can explain simultaneously the generation of the
baryon and the dark matter in the Universe, and explain their
surprisingly close values; (v) AD model, due to its high efficiency,
can be successful even in case of significant production of entropy
at late times, predicted by some particle physics models.

AD scenario is based on SUSY. In supersymmetric models  scalar
superpartner of baryons and leptons $\varphi$ exist. The potential
U($\varphi$) of such scalar field may have flat directions, along
which the field can have a non-zero vacuum expectation value due to
quantum fluctuations during inflation. After inflation $\varphi$
evolves down to the equilibrium point $\varphi$ = 0 and if the
potential U($\varphi$) is not symmetric with respect to the phase
rotation it acquires non-vanishing and typically large baryon
charge. Subsequent B-conserving decay of $\varphi$ into quarks and
leptons transform baryon asymmetry into the quarks sector. In
contrast to other scenarios of baryogenesis, in which the generated
asymmetry usually is insufficient, the original  Afleck-Dine
scenario leads to higher value of $\beta$ and additional mechanisms
are needed to dilute it down to the observed value.

AD mechanism was re-examined in refs.~\cite{AD2}. It was realized
that finite energy density of the early Universe breaks SUSY  and
induces soft parameters in the soft potential along flat directions,
which are of the order of the Hubble parameter.  Then, contrary to
the original AD mechanism the observed value of the baryon asymmetry
may be generated, without the requirement of subsequent entropy
release. Different issues on AD baryogenesis were presented in
refs.~\cite{condT,anis,allaverdi}. AD baryogenesis mechanism was
used in numerous SM extensions and different inflationary scenario.
Just to list several of the more recent studies: AD in effective
supergravity~\cite{ADsugra}, AD in anomaly mediated SUSY breaking
models~\cite{ADanomal}, AD in SUSY with R-parity
violation~\cite{ADR}, AD in D-term inflation~\cite{ADinf}, etc. Most
of AD baryogenesis models can naturally explain the origin of the
dark matter in the Universe. Constraints on the sub-class of AD
models was obtained from current CMB data, based on the backreaction
of the flat direction on the inflationary potential~\cite{ADc}.

Here we discuss the scalar field condensate baryogenesis model (SFC
baryogenesis), which is among the preferred today baryogenesis
scenarios, compatible with inflation. It is based on the
Affleck-Dine scenario.

 SFC baryogenesis model  was first
discussed and studied analytically in refs.\cite{dolgkiril90,
dolgkiril91}. There it was shown that the account of particle
creation by the time varying scalar field during post-inflationary
period will lead to strong reduction of the produced baryon excess
in the Affleck-Dine scenario. Namely,  it was proven that fast
oscillations of $\varphi$ result in particle creation due to the
coupling of the scalar field to fermions $g\varphi f_1f_2$ , where
$g^2/4\pi = \alpha$.  For $\lambda_i^{3/4}
> g$ the rate of particle creation $\Gamma$ exceeds the ordinary decay rate
of $\varphi$
 at the stage of baryon non-conservation and, therefore, its  amplitude is damped.
Hence,  the baryon charge, contained in the condensate, is reduced
due to particle creation at this stage with considerable baryon
violation. \footnote{In case of $\Gamma=const$ the baryon charge,
contained in the condensate, is reduced exponentially and it does
not survive till $\varphi$ decays to quarks and leptons. In case
when $\Gamma$ is a decreasing function of time the damping process
may be slow enough for the baryon charge contained in $\varphi$ to
survive until the B-conservation epoch.}

The importance of a precise numerical account for the particle
creation processes was explored further in
refs.\cite{chizkiril96,kirilpanay07}.
 Different possibilities of SFC baryogenesis models were
discussed. The possibility to generate simultaneously, within
inhomogeneous SCF baryogenesis model, the observed baryon asymmetry
and the observed large scale structure quasi-periodicity of the
baryonic matter was studied in
refs.\cite{mih,kirilchiz00a,chizkiril96}. On the basis of
inhomogeneous SCF baryogenesis elegant mechanisms  were proposed for
achieving sufficient separation between domains of matter and
antimatter (to inhibit the contact and evade annihilation of matter
and antimatter regions with big density), that allow the production
of considerable antimatter domains with different size in the
Universe
 and their observational signatures were analyzed~\cite{kawa,kirilpanay02,mih,NPS,anti1,rev2}.

 In series
of papers \cite{kirilpanay07,kirilpanay14,kirilpanay12}, we explored
numerically SFC baryogenesis model. Here we present the results of
our extended numerical analysis of the evolution of the baryon
excess in SFC baryogenesis model and its dependence on the model
parameters.

In the next section we briefly describe the SCF baryogenesis model
and the numerical approach we have used. The last section presents
the results, i.e. we present the value of the produced baryon
density for numerous sets of model's parameters.

%------------------------------------------------

\section{SFC baryogenesis model}
\subsection{Description}
The essential ingredient of the model is a baryon charged complex
scalar field $\varphi$,  present  together with the inflaton.  A
condensate $<\varphi>\neq0$ with a nonzero baryon charge is formed
during the inflationary period as a result of the rise of quantum
fluctuations of the $\varphi$ field \cite{vilen82, linde82, bunch78,
star82}: $<\varphi^2>=H^3t/4\pi^2$ untill the limiting value
$<\varphi^2> \sim H^2/ \sqrt {\lambda}$ in case that $\lambda
\varphi^4$ terms dominate in the potential energy of $\varphi$.

The baryon charge of the field is not conserved at large field
amplitude due to the presence of the $B$ nonconserving
self-interaction terms in its potential.

We choose the form of the potential as follows:

\begin{equation} U(\varphi)=m^2\varphi^2+{\lambda_1\over 2}|\varphi|^4+
{\lambda_2\over 4}(\varphi^4+\varphi^{*4})+ {\lambda_3\over
4}|\varphi|^2(\varphi^2+\varphi^{*2}) \end{equation}

The mass parameters of the potential are assumed small in comparison
with the Hubble constant during inflation $m \ll  H_I$.  In
supersymmetric theories the self coupling constants $\lambda_i$ are
of the order of the gauge coupling constant $\alpha$. A natural
range of $m$ is $10^2 - 10^4$ GeV.

We examine the case when after inflation there exist two scalar
fields - the inflaton $\psi$ and the scalar field $\varphi$ and the
inflaton density dominates prior the decay of $\varphi$:
$\rho_{\psi}>\rho_{\varphi}$. Hence, at the end of inflation the
Hubble parameter is $H=2/(3t)$.

In the expanding Universe, in case of spatially homogeneous field
$\varphi$ satisfies the equation of motion:

\begin{equation} \ddot{\varphi}+3H\dot{\varphi}+
{1 \over4} \Gamma\dot{\varphi}+ U'_{\varphi}=0,
\end{equation}

where a(t) is the scale factor and $H=\dot{a}/a$, $\Gamma$ accounts
for the particle creation processes.

The initial values for the field variables are derived from the
natural assumption that the energy density of $\varphi$ at the
inflationary stage is of the order $H^4_I$, then

\begin{equation} \varphi^{max}_o \sim H_I\lambda^{-1/4} and \quad \dot{\varphi_o}=(H_I)^2.
\end{equation}

After inflation $\varphi$ oscillates around its equilibrium point
and its amplitude decreases due to the Universe expansion and the
particle creation by the oscillating scalar field. In case $\Gamma$
is a decreasing function of time the damping process may be slow
enough for the baryon charge contained in $\varphi$ to survive until
the B-conservation epoch \cite{dolgkiril91}.

At low $\varphi$ baryon violation (BV) becomes negligible. At the B
conserving stage the baryon charge contained in the field is
transferred to that of quarks during the decay of the field
$\varphi\to q\bar{q} l \gamma$ at $t_b$. As a result, in case
$\varphi$ has not reached the equilibrium point at $t_b$, the
baryogenesis makes a snapshot of $\varphi(t_b)$ and a baryon
asymmetric plasma appears. This asymmetry, eventually further
diluted during the following evolution of the Universe, gives the
present observed baryon asymmetry of the Universe.

\subsection{Evolution of the baryon charge carrying field}

We have solved the system of ordinary differential equations,
corresponding to the equation of motion for the real and imaginary
components  of $\varphi = x + iy$:

\begin{eqnarray}
&&\ddot{x}+3H\dot{x}+{1 \over 4} \Gamma_x \dot{x}+
(\lambda+\lambda_3)x^3+\lambda'xy^2=0 \nonumber
\\
&&\ddot{y}+3H\dot{y}+{1 \over 4} \Gamma_y \dot{y}+
(\lambda-\lambda_3)y^3+\lambda'yx^2=0 \label{motion}
\end{eqnarray}
where $\lambda=\lambda_1+\lambda_2$,
$\lambda'=\lambda_1-3\lambda_2$.

It is convenient to make the substitutions
$x=H_I(t_i/t)^{2/3}u(\eta)$, $y=H_I(t_i/t)^{2/3}v(\eta)$ where
$\eta=2(t/t_i)^{1/3}$. Then the functions $u(\eta)$ and $v(\eta)$
satisfy the equations:

\begin{eqnarray}
&& u''+ 0.75\; \alpha\Omega_u (u'-2u\eta^{-1})+
u[(\lambda+\lambda_3) u^2+\lambda'v^2-2\eta^{-2}+ {m \over H}^2
\eta^4]=0 \nonumber
\\
&& v''+ 0.75\; \alpha\Omega_v (v'-2v\eta^{-1})+
v[(\lambda-\lambda_3) v^2+\lambda'u^2-2\eta^{-2}+ {m \over H}^2
\eta^4]=0. \label{urav}
\end{eqnarray}

%The initial conditions in the new variables are:

%\begin{eqnarray}
%&& u_0(\eta) = \lambda^{-0.25}2^{0.25}\cos{p}, \quad v_0(\eta) =
%\lambda^{-0.25}2^{0.25}\sin{p} \quad and \nonumber
%\\
%&& u'_0=3/2^{1.5}+ \lambda^{-0.25}2^{0.25}\cos{p}, \quad
%v'_0=3/2^{1.5}+ \lambda^{-0.25}2^{0.25}\sin{p}.
%\end{eqnarray}

%where p is the angle.

The baryon charge in the comoving volume $V=V_i(t/t_i)^2$ is given
by

\begin{equation}
B=N_B \cdot V=2 (u'v-v'u).
\end{equation}

%\subsection{Numerical analysis of the scalar field evolution and the
%baryon charge contained in it}
We have  solved numerically  the system of ordinary differential
equations (\ref{urav}), corresponding to the equation of motion for
the real and imaginary part of $\varphi$ and B contained in it,
using Runge-Kutta 4th order method and fortran 77. The Runge-Kutta
4th order routine from \cite{numrec} is used.

We studied numerically the evolution of $\varphi(\eta)$ and
$B(\eta)$ in the period after inflation until the BC epoch. The
typical range of energies discussed was $10^{15} - 100$ GeV.
Therefore, serious computational resources were used. A single
calculation took between several hours and three weeks, depending on
the concrete parameters.

We analyzed $\varphi$ and B evolution for natural ranges of values
of the model's parameters: $\lambda = 10^{-2} - 5 \times 10^{-2}$,
$\alpha = 10^{-3} - 5 \times 10^{-2}$,  $H_I = 10^7 - 10^{16}$ GeV,
$m = 100 - 1000$ GeV. The numerical analysis was provided for around
seventy sets of parameters.

We have accounted numerically for the particle creation processes by
varying $\varphi$, which allowed to describe more precisely the
evolution of $B$ and determine its final value which was transferred
to quarks (antiquarks) at $t_b$ epoch and defined the baryon
asymmetry. In this work we calculated $\Gamma$ numerically in
contrast to our previous papers, where for the rate of particle
creation $\Gamma$ the analytical estimation was used $\Gamma=\alpha
\Omega$, where $\Omega \sim \lambda_1^{1/4} \varphi$. In the program
$\Omega_u$ and $\Omega_v$ ware calculated at each step in separate
routine procedures.

The results of our numerical study are presented in the next
section.

\section{The generated baryon charge for different parameters values.
Numerical results.}

We have calculated $B$ for different sets of values of model's
parameters - gauge coupling constant $\alpha$, Hubble constant
during inflation $H_i$, mass of the condensate $m$ and self coupling
constants $\lambda_i$. We have not made calculations for all the
possible values of the parameters, because each single point
requires days or weeks of CP time. Our main aim was to find the
dependence of the final B on the parameters  and choose the more
promising ranges of the parameters for successful baryogenesis,
rather than providing full systematic numerical study. Therefore,
some entries in the tables below are missing.

\subsection{Dependence on Hubble constant during inflation $H_I$}

We have followed the evolution $B(\eta)$ varying $H_I$ for fixed
values of the other parameters. The results of our preliminary
analysis of this dependence  have been first discussed in ref.
\cite{kirilpanay07,kirilpanay12}.

In this work we have performed an extended  analysis of this
dependence, studying wider range of models parameters.

\begin{table}[h]
\begin{center}
%\begin{adjustbox}{max width=\textwidth}
\begin{tabular}{|lr|r|r|r|r|r|r|r|}
  \hline
   & $H$, GeV & \verb"1.00E+14" & \verb"1.00E+12" & \verb"1.00E+11"
    & \verb"1.00E+10" & \verb"1.00E+09" & \verb"1.00E+08" & \verb"1.00E+07"  \\
   $m$, GeV & & & & & & & & \\
  \hline
   350 & & \verb"-2.75E-07" & \verb"2.10E-06" & \verb"1.18E-04" &
   \verb"1.61E-03" & \verb"2.89E-02" & \verb"-4.08E-01" & \verb"-2.31E-01" \\
  \hline
   500 & & & \verb"7.44E-05" & \verb"-7.72E-04" & \verb"-6.72E-04"
    & \verb"-1.00E-02" & \verb"2.69E-01" & \verb"1.58E-01" \\
  \hline
   800 & & & \verb"-1.07E-04" & \verb"-1.40E-03" & \verb"2.27E-03"
    & \verb"2.86E-02" & \verb"1.11E-01" & \verb"9.04E-01" \\
  \hline
\end{tabular}
%\end{adjustbox}
  \caption{\label{B_H_a01_l101_l23001} The baryon charge
$B$ contained in the SFC at the time of its decay for different
$H_I$ and $m$ and fixed set of $\lambda_1= 10^{-2}$, $\alpha =
10^{-2}$ and $\lambda_2=\lambda_3=10^{-3}$ -
$\varphi_o=H_I\lambda^{-1/4}$ and $\dot{\varphi}_o={H_I}^2$. The
particle creation processes are accounted for numerically.}
\end{center}
\end{table}

%\begin{figure}
%\centering \includegraphics[width=10 cm]{B_H_a01_l101_l23001.eps}
%\caption{\textbf{The evolution of the baryon charge
%$B(\eta)$ for different $H_I$ and m and fixed set of $\lambda_1= 10^{-2}$, $\alpha = 10^{-2}$ and $\lambda_2=\lambda_3=10^{-3}$} - $\varphi_o=H_I\lambda^{-1/4}$ and $\dot{\varphi}_o={H_I}^2$. The particle creation processes are accounted for
%numerically.} \label{B_H_a01_l101_l23001}
%\end{figure}

In the rows in Tabl. \ref{B_H_a01_l101_l23001},  Tabl.
\ref{B_H_a01_l105_l23001} and Tabl. \ref{B_a_l105_l2l3001_m350}
 we present the results of
the generated baryon charge in the SFC baryogenesis model for
different values  of the Hubble constant $H_I$ and for different
fixed values of the other parameters of the model. The first table
presents $B(H_I)$ for $\lambda_1 = 10^{-2}$, $\alpha = 10^{-2}$,
$\lambda_2=\lambda_3 = 10^{-3}$, the second one presents it for
$\lambda_1 = 5 \times 10^{-2}$, $\alpha = 10^{-2}$,
$\lambda_2=\lambda_3 = 10^{-3}$, the third one presents it for
$\lambda_1= 5 \times 10^{-2}$, $\lambda_2=\lambda_3=10^{-3}$,
$m=350$ GeV and two different values of  $\alpha$, namely  $10^{-2}$
and $5.10^{-2}$.

\begin{table}[h]
\begin{center}
%\begin{adjustbox}{max width=\textwidth}
\begin{tabular}{|lr|r|r|r|r|r|r|}
  \hline
   & $H$, GeV & \verb"1.00E+12" & \verb"1.00E+11"
    & \verb"1.00E+10" & \verb"1.00E+09" & \verb"1.00E+08" & \verb"1.00E+07"  \\
   $m$, GeV & & & & & & & \\
  \hline
   350 & & \verb"1.00E-05" & \verb"1.00E-04" &
   \verb"-5.00E-05" & \verb"-1.00E-04" & \verb"7.00E-03" & \verb"-4.00E-03" \\
  \hline
   500 & & \verb"2.57E-06" & \verb"1.00E-06" & \verb"-2.00E-04"
    & \verb"-1.34E-02" & \verb"7.87E-02" & \verb"5.00E-01" \\
  \hline
\end{tabular}
%\end{adjustbox}
  \caption{\label{B_H_a01_l105_l23001}The  baryon charge
$B$ contained in the SFC at the time of its decay for different
$H_I$ and m and fixed set of $\lambda_1= 5 \times 10^{-2}$, $\alpha
= 10^{-2}$ and $\lambda_2=\lambda_3=10^{-3}$ -
$\varphi_o=H_I\lambda^{-1/4}$ and $\dot{\varphi}_o={H_I}^2$.}
\end{center}
\end{table}

\begin{figure}[!h]
\centering \includegraphics[width=15 cm]{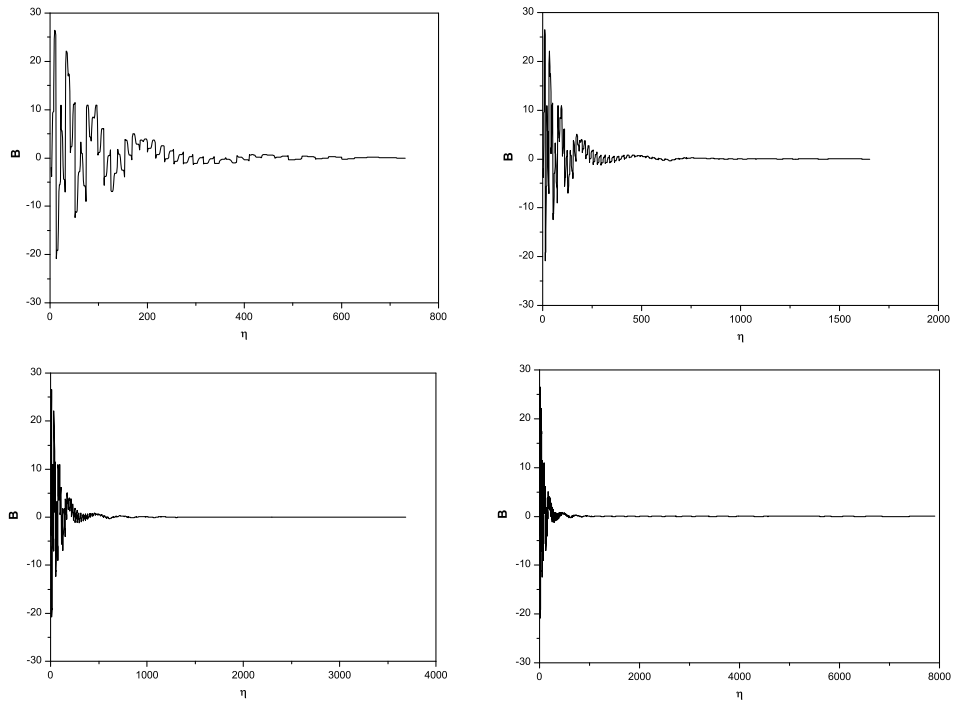}
\caption{The evolution of the baryon charge $B(\eta)$ for
$\lambda_1= 10^{-2}$, $\lambda_2=\lambda_3=10^{-3}$,
$\alpha=10^{-2}$, $m=500$ GeV, $\varphi_o=H_I\lambda^{-1/4}$ and
$\dot{\varphi}_o={H_I}^2$. The upper left plot is for $H_I=10^{9}$
GeV, the upper right plot is for $H_I=10^{10}$ GeV, the lower left
plot is for $H_I=10^{11}$ GeV and the lower right plot is for
$H_I=10^{12}$ GeV.} \label{m500}
\end{figure}

%\begin{figure}
%\centering \includegraphics[width=10 cm]{B_H_a01_l105_l23001.eps}
%\caption{\textbf{The evolution of the baryon charge
%$B(\eta)$ for different $H_I$ and m and fixed set of $\lambda_1= 5 \times 10^{-2}$, $\alpha = 10^{-2}$ and $\lambda_2=\lambda_3=10^{-3}$} - $\varphi_o=H_I\lambda^{-1/4}$ and $\dot{\varphi}_o={H_I}^2$. The particle creation processes are accounted for
%numerically.} \label{B_H_a01_l105_l23001}
%\end{figure}
 Fig. \ref{m500} presents the dependence of the evolution of the  baryon
charge on  $H_I$ at fixed values of the other parameters.

Our detail analysis for numerous different parameters of the SCF
model shows that
 {\it B evolution becomes longer and the final B
value decreases with the increase of $H_I$.} This result is in agreement with
previous numerical and analytical  studies.  It is an expected
result because particle creation, which reduces $\beta$ is
proportional to $\varphi$, $\Gamma \sim \Omega \sim \varphi$, and
the initial value of $\varphi$ is proportional to $H_I$. Thus, the
bigger $H_I$ - more efficient is the decrease of $\beta$ due to
particle creation.

\subsection{Dependence on gauge coupling constant $\alpha$}

Using the numerical account for $\Gamma$ we have calculated
$B(\eta)$ for $\alpha$ varying in the range $10^{-3} - 10^{-2}$ and
fixed other parameters, see also refs.~\cite{kirilpanay07}. The
dependence of $B$ on $\alpha$ is very strong, as can be expected,
knowing that particle creation processes play essential role for the
evolution of the field and the baryon charge, contained in it, and
keeping in mind that the analytical estimation is $\Gamma=\alpha
\Omega$.

\begin{table}[h]
\begin{center}
%\begin{adjustbox}{max width=\textwidth}
\begin{tabular}{|lr|r|r|r|r|r|r|}
  \hline
   & $H$, GeV & \verb"1.00E+12" & \verb"1.00E+11"
    & \verb"1.00E+10" & \verb"1.00E+09" & \verb"1.00E+08" & \verb"1.00E+07"  \\
   $\alpha$ & & & & & & & \\
  \hline
   \verb"1.00E-02" & & \verb"1.00E-05" & \verb"1.00E-04" &
   \verb"-5.00E-05" & \verb"-1.00E-04" & \verb"7.00E-03" & \verb"-4.00E-03" \\
  \hline
   \verb"5.00E-02" & & \verb"8.01E-07" &  & \verb"5.50E-05"
    & \verb"2.40E-03" & \verb"8.10E-03" & \verb"6.90E-02" \\
  \hline
\end{tabular}
%\end{adjustbox}
  \caption{\label{B_a_l105_l2l3001_m350}The baryon charge
$B$ contained in the SFC at the time of its decay for different
$\alpha$ and $H_I$ and $\lambda_1= 5 \times 10^{-2}$,
$\lambda_2=\lambda_3=10^{-3}$, $m=350$ GeV -
$\varphi_o=H_I\lambda^{-1/4}$ and $\dot{\varphi}_o={H_I}^2$.}
\end{center}
\end{table}

\begin{figure}[!h]
\centering \includegraphics[width=8 cm]{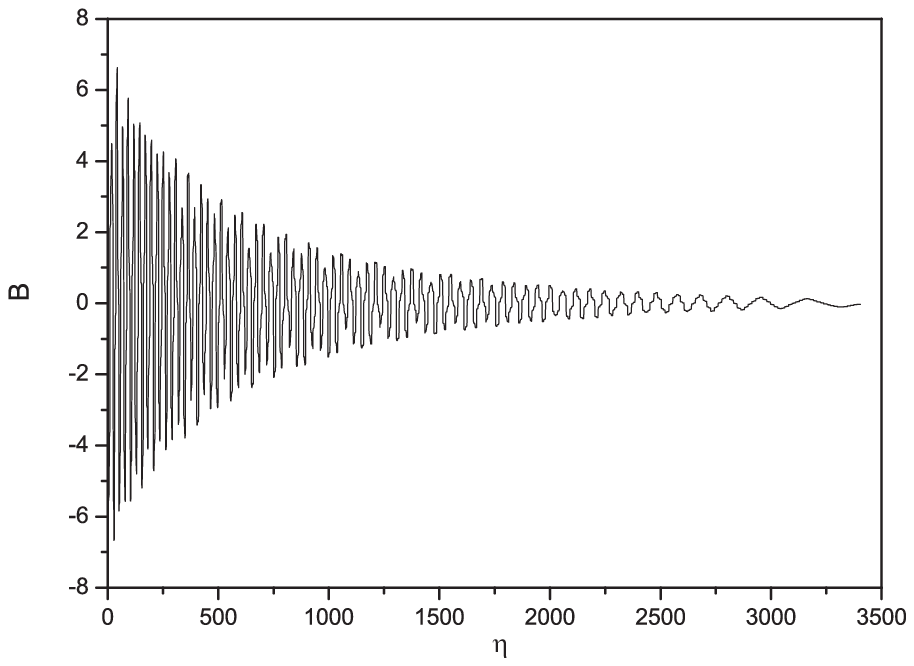} \centering
\includegraphics[width=8 cm]{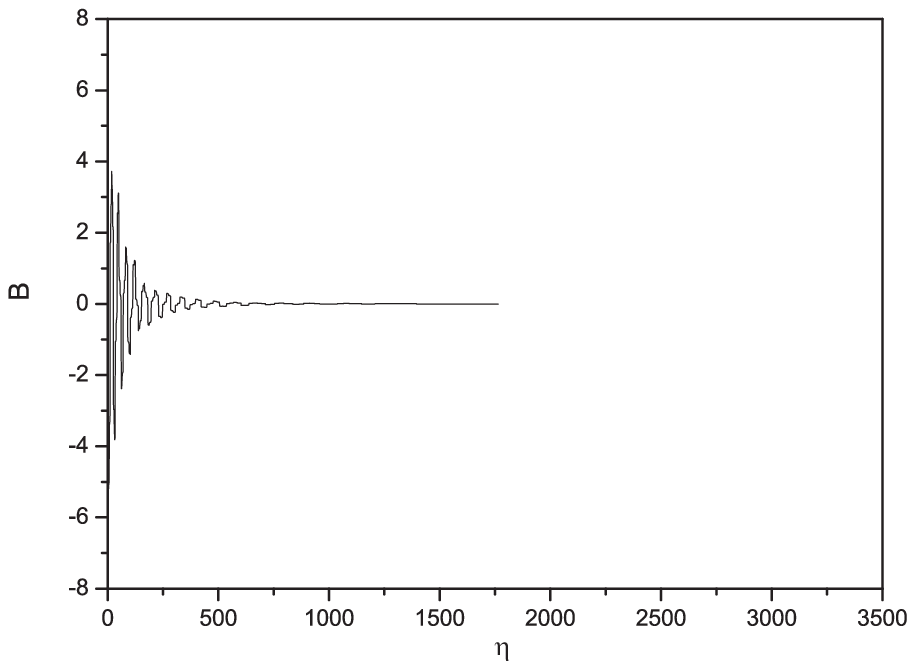} \centering
\includegraphics[width=8 cm]{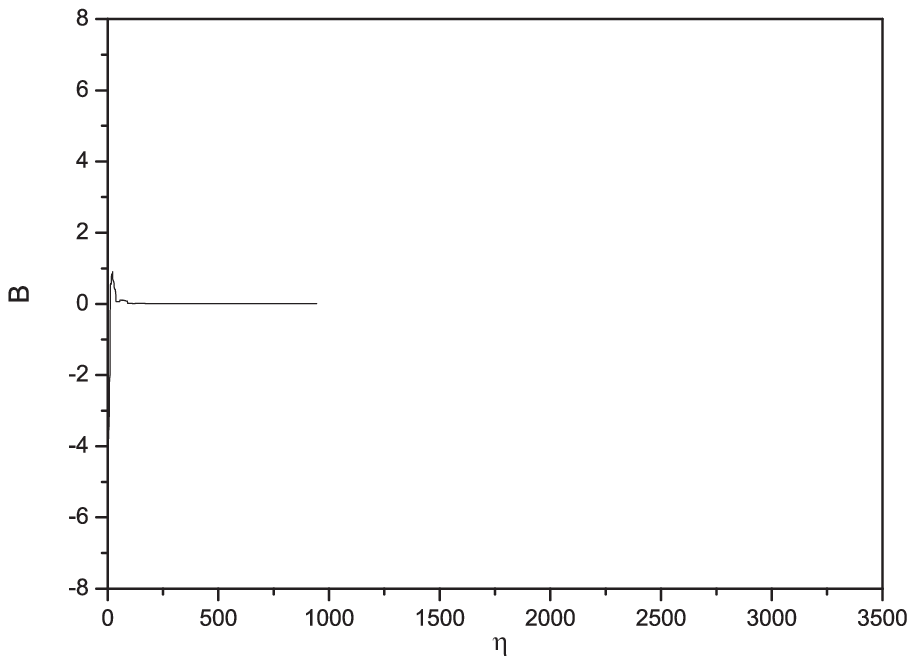}
 \caption{The
evolution of the baryon charge $B(\eta)$ for $\lambda_1=5 \times
10^{-2}$, $\lambda_2=\lambda_3= 10^{-3}$, $H=10^{10}$ GeV, $m= 350$
GeV, $\varphi_o=H_I\lambda^{-1/4}$ and $\dot{\varphi}_o={H_I}^2$.
The upper left plot is for $\alpha=10^{-3}$, the upper right plot is
for $\alpha=10^{-2}$, the bottom plot is for $\alpha=5 \times
10^{-2}$.} \label{alpha}
\end{figure}

{\it With increasing $\alpha$, $B$ evolution becomes shorter and the
final $B$ decreases.} An illustration of this dependence $B(\alpha)$
is given in Fig. \ref{alpha}.

In the columns of Tabl. \ref{B_a_l105_l2l3001_m350} we present the
results for the baryon charge contained in the SFC at the time of
its decay for different values of $\alpha$ and for $\lambda_1 = 5
\times 10^{-2}$, $\lambda_2=\lambda_3 = 10^{-3}$, $m = 350$ GeV.

\subsection{Dependence on the mass $m$ of the condensate}

 The dependence of the final baryon charge on
 $m$ for fixed $\lambda_1$, $\lambda_2$, $\lambda_3$, $\alpha$ and
$H_I$  has been first discussed in ref.
\cite{kirilpanay07,kirilpanay12}. Here we present the results of a
more detail study. In table 2 the dependence of generated baryon
asymmetry values on the mass of the field is given for different
fixed sets of the other parameters of the model.

\begin{figure}[!h]
\centering \includegraphics[width=8 cm]{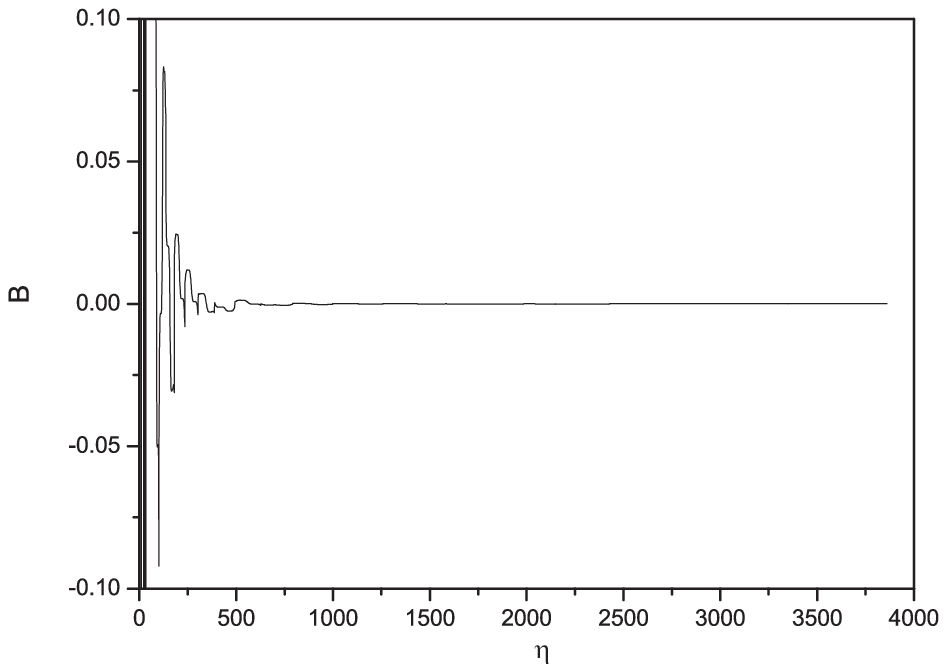} \centering
\includegraphics[width=8 cm]{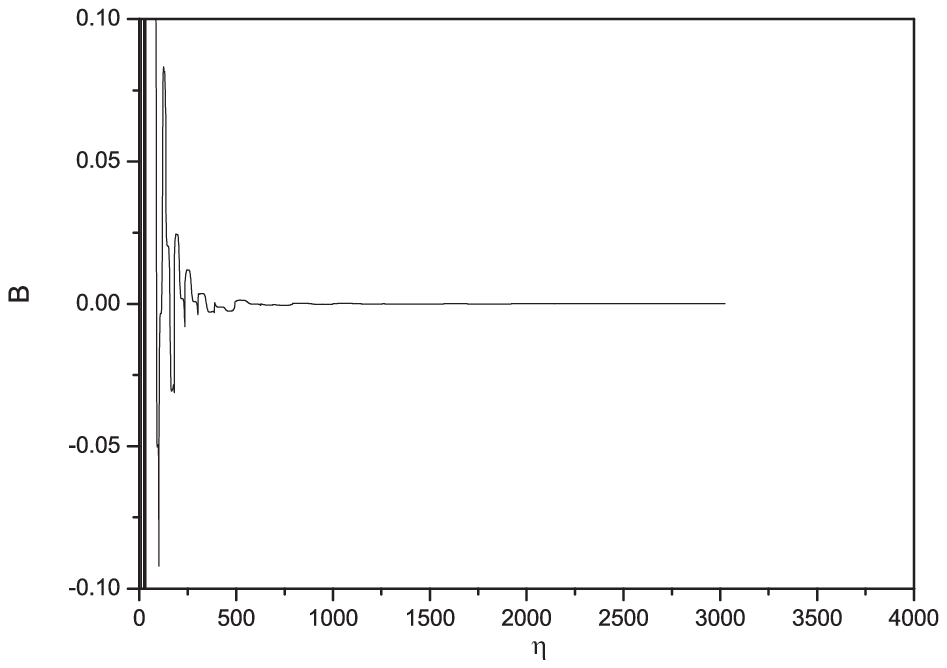} \centering
\includegraphics[width=8 cm]{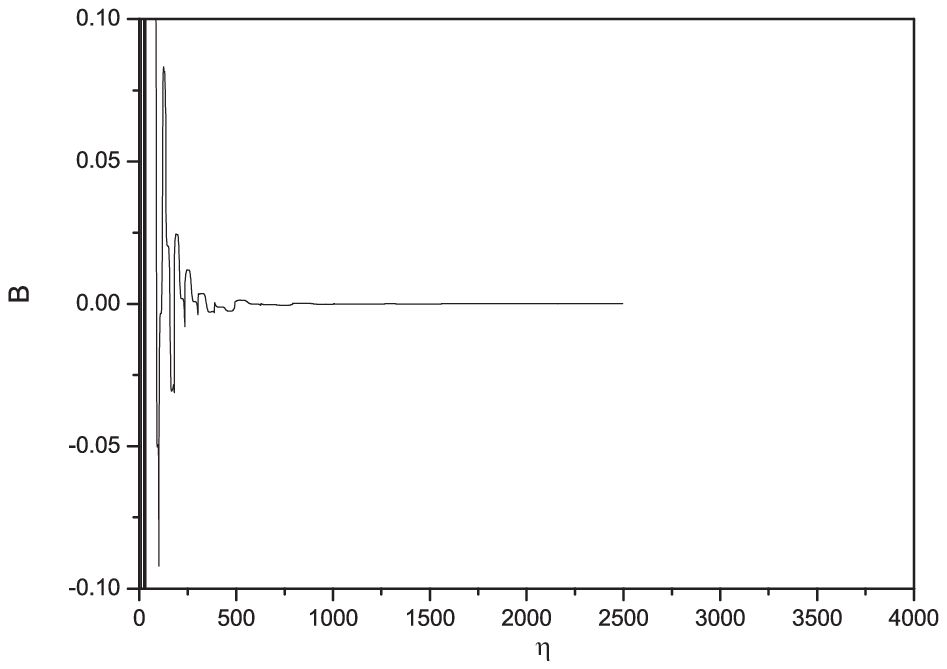}
\caption{The evolution of the baryon charge $B(\eta)$ for
$\lambda_1=5 \times 10^{-2}$, $\lambda_2=\lambda_3= 10^{-2}$,
$\alpha=5 \times 10^{-2}$, $H=10^{11}$ GeV,
$\varphi_o=H_I\lambda^{-1/4}$ and $\dot{\varphi}_o={H_I}^2$. The
upper left plot curve is for $m = 100$ GeV, the upper right plot is
for $m = 200$ GeV, the bottom plot is for $m = 350$ GeV.} \label{mm}
\end{figure}

It has been found that in general $B$ decreases with the increase of
the value of $m$.  This behavior is more clearly and more strongly
expressed for big values of $H_I$  and then corresponds to the
expected one from analytical estimations (Namely,  as far as $m$
defines the onset of BC epoch: $t_{b} \sim 1/ \alpha m$, and hence
for lower values of $m$, $B$ evolution is longer and the final $B$
value is smaller.) The dependence is illustrated in Fig. \ref{mm}.

For smaller values of $H_I$ the dependence is weaker and not so
straightforward and clear. In table 2 we present the generated
baryon asymmetry values for different mass of the field and fixed
values of the other parameters of the model.

In the columns of Tabl. \ref{B_H_a01_l101_l23001} and Tabl.
\ref{B_H_a01_l105_l23001} we present the results for the baryon
charge contained in the SFC at the time of its decay for different
values  of the mass of the field $m$ and for different fixed values
of the other parameters of the model.

\subsection{Dependence on the self-coupling constants}

The dependence of the baryon charge, at the B-conservation epoch, on
the value of the coupling constants $\lambda_i$ was first discussed
in ref. \cite{kirilpanay14}. Our extended analysis confirms that the
final B value decreases when increasing $\lambda_1$ and  $B$
evolution becomes shorter. The effect provides  a difference in the
final $B$ value of an order of magnitude.

The results of the extended analysis showed that the final value of
$B$ may be much sensitive to  $\lambda_2$ and $\lambda_3$ than
previously estimated. Namely, we have found that the final  values
of $B$ may differ up to 3 orders of magnitude even for small changes
of these parameters. For example for $m=350$ GeV, $H_I=10^{10}$,
$\alpha=10^{-3}$, $\lambda_1=5 \times 10^{-2}$ the final B value for
$\lambda_2=\lambda_3=5 \times 10^{-4}$ is $B=3.88 \times 10^{-5}$,
while for $\lambda_2=\lambda_3= 10^{-3}$ is $B=2.36 \times 10^{-2}$.

Thus, we have found that the dependence of baryon generation on the
self-coupling constants is  important for determination of the
parameters range for the successful baryogenesis model.

\section{Estimation of the generated baryon asymmetry}

In order to estimate the baryon asymmetry on the basis of the
obtained results for the produced baryon density it is necessary to
know the temperature of the relativistic plasma after the decay of
$\varphi$ and the decay of the inflaton.

In case the inflaton energy density dominates until the decay of
$\varphi$, i.e. prior to reheating, $\rho_{\psi}>\rho_{\varphi}$,
the entropy is mainly defined by the relativistic particles from
inflaton decay. Thus, the temperature after the decay of $\psi$ at
$t_{\psi}$ will be approximately
\begin{equation}
T_R \sim (\rho_{\psi})^{1/4} =
(\rho^0_{\psi})^{1/4}(\eta_0/\eta_{\psi})^{3/2}.
\end{equation}

Then the baryon asymmetry will be given by:

%\begin{equation}
% \beta \sim N_B/T_R^3 \sim (B/H_I) (M_{Pl}/t_{\psi})^{1/2}
%\end{equation}
\begin{equation}
 \beta \sim N_B/T_R^3 \sim B T_R/H_I
\end{equation}

where $T_R$ is the reheating temperature after the decay of the
inflaton.

Hence, from these estimations it is seen that the lower the
reheating temperature after inflaton decay the lower the produced
baryon asymmetry will be. Also, the later the inflaton decays the
smaller the produced $\beta$ will be. Knowing that the reheating
temperature should be sufficiently low to avoid gravitino problem,
i.e. in our model it should be several orders or more lower than the
value of $H_I$, and having the results for $B$, it is easy to obtain
the value of the observed baryon asymmetry for different sets of
parameters in this model.

%For example, substituting $T_R=10^9$ GeV and $H_I=10^{12}$ GeV and
%$B=8. 10^{-7}$ in the above formula  (see Table 3, first column for
%$\alpha=5.10^{-2}$) the produced baryon asymmetry is similar to the
%observed one.

Of course the $H_I$, the decay time of $\psi$ and the value of the
reheating temperature may be different in different inflationary
scenarios.  We would like only to note here that the results of the
numerical analysis of the SCF baryogenesis model are encouraging.
The analysis points that this model provides an opportunity to
produce baryon asymmetry $\beta$, consistent with its observed value
for  natural values of the model's parameters. Therefore, this model
deserves further considerations.

\section{Conclusions}

It was found that the  analytical estimations of the baryon charge
evolution and its final value in SFC baryogenesis model may
considerably differ from the exact numerically calculated ones.
Therefore, in this work we have numerically explored the SFC
baryogenesis model for numerous sets of model's parameters.

We have investigated the dependence of the evolution of the field
and the evolution of the baryon charge contained in it, as well as
  the final value of the baryon charge contained in it  on the
model's parameters: the gauge coupling constant $\alpha$, the Hubble
parameter at the inflationary stage $H_I$, the mass m and the
self-coupling constants $\lambda_i$. Qualitative dependence of the
final B on these parameters have been found. Namely, it was shown
that the produced baryon excess is a strongly decreasing function of
$\alpha$, it also is a decreasing function of $H_I$. The dependence
on $m$ is not so straightforward. For small $m$ values $B$ decreases
with $m$ increase, however for larger $m$  the dependence is more
complicated.

 The  analysis may be used to  indicate the  values of
 the model's parameters for which
 baryon asymmetry $\beta$, consistent with its observed
value, may be produced in a given inflationary scenario.

The results of this analysis may be used for constructing realistic
SCF  baryogenesis models. Moreover, assuming  SCM baryogenesis and
assuming a concrete inflationary scenario, from the observed value
of the baryon asymmetry it is possible to put cosmological
constraints on the model's parameters, provided by physics theories,
i. e. constrain physics beyond Standard model.

\section*{Acknowledgements}
We would like to thank M. V. Chizhov for useful conversations and
for the technical support. D.K. expresses her gratitude to O. V.
Teryaev for the overall help during her visiting position at BLTP,
JINR. The authors are grateful to the unknown referee for the useful
suggestions and criticism.

%----------------------------------------------------------------------------------------

%\end{multicols}

\end{document}